\documentclass[showpacs, twocolumn]{revtex4-1}
\usepackage{graphicx}
\usepackage{amsmath}
\usepackage{appendix}

\begin{document}

\title{Many-body and temperature effects in two-dimensional quantum droplets in Bose-Bose mixtures}

\author{ Abdel\^{a}ali Boudjem\^{a}a $^{1,2}$ }
\affiliation{
$^1$ Department of Physics, Faculty of Exact Sciences and Informatics, Hassiba Benbouali University of Chlef, P.O. Box 78, 02000, Ouled-Fares, Chlef, Algeria.\\
$^2$Laboratory of Mechanics and Energy, Hassiba Benbouali University of Chlef, P.O. Box 78, 02000, Ouled-Fares, Chlef, Algeria.}
\email {a.boudjemaa@univ-chlef.dz}


\begin{abstract}

We study the equilibrium properties of self-bound droplets in two-dimensional Bose mixtures employing the time-dependent Hartree-Fock-Bogoliubov theory. 
This theory allows one to understand both the many-body and temperature effects beyond the Lee-Huang-Yang description. 
We calculate higher-order corrections to the excitations, the sound velocity, and the energy of the droplet. 
Our results for the ground-state energy are compared with the diffusion Monte Carlo data and good agreement is found.
The behavior of the depletion and anomalous density of the droplet is also discussed.
At finite temperature, we show that the droplet emerges at temperatures well below the Berezinskii-Kosterlitz-Thouless transition temperature.
The critical temperature strongly depends on the interspecies interactions.
Our study is extended to the finite size droplet by numerically solving the generalized finite-temperature Gross-Pitaevskii equation which is obtained self-consistently from our formalism
in the framework of the local density approximation.
\end{abstract}

\maketitle

\section*{Introduction}

Recently, the investigation of self-bound droplet states in Bose mixtures \cite{Petrov, Cab, Cab1, Sem,Err} and dipolar ultracold gases \cite{Pfau,Pfau1,Chom}
has become a burgeoning area of interest.
This novel state of matter forms due to the intriguing competition  between the attractive mean-field interactions 
and the repulsive force furnished by the Lee-Huang-Yang (LHY) quantum fluctuations.
The most important feature of these peculiar breakthroughs is that they are ultradilute contrary to the liquid Helium droplets \cite{Vol}.
Quantum droplets have been extensively researched in various contexts (see for review \cite {Luo,Pfau2,Guo} and references therein).

The binary  Bose-Einstein condensates (BECs) with weak attractive inter- and repulsive intraspecies interactions 
support also the creation of ultradilute liquid-like droplets in two-dimensional (2D) configuration \cite{Astr}.
In the last few years, 2D quantum droplets have been studied from a variety of aspects including:
effects of the spin-orbit coupling \cite{Li}, formation and stability of quantum Rabi-coupled  droplets \cite{Chiq},  
superfluidity and vortices  \cite{Yong},  supersolid stripe phase \cite{Sach}, dynamical excitations \cite{Abdullaev}, bulk properties and quantum phases \cite{Hu,Pan}.

In previous studies, the most commonly employed  theoretical tool for describing the static and the dynamics of the droplet is
the generalized Gross-Pitaevskii equation (GPE). 
Although this model which is based on the Petrov's theory \cite{Astr} gives reasonable results, it suffers from different handicaps.
First, the generalized GPE disagrees with some experimental measurements \cite{Cab} and quantum Monte Carlo (QMC) method \cite{Cik,Ota}. 
In addition, it fails to properly predict the critical atom number \cite{Pan,Boudj1} and to describe effects of quantum correlations \cite{Boudj1}. 
In this regard, many theoretical works beyond the generalized GPE have been introduced  to study the properties of self-bound droplets \cite{Hu, Pan,Cik,Ota, Cap,Cap1,Stau,Gu,Cik1}.
Among them, the pairing theory \cite{Hu} which has been used in order to improve the Petrov's theory \cite{Astr} for quantum droplets of 2D Bose mixtures.
However, the pairing approach gives almost the same results as the Petrov's theory for the ground-state energy. 
Both theories diverge from the diffusion Monte Carlo (DMC) simulation notably in the regime of small interspecies attraction.
This discrepancy can be attributed to the absence of higher-order corrections that are crucial in 2D Bose systems.

Very recently, we have developed an interesting theoretical model beyond the standard LHY  \cite{ Boudj1, Boudj2, Boudj3,Boudj18} 
called the time-dependent Hartree-Fock-Bogoliubov theory (TDHFB) able to selfconsistently explain the behavior of quantum self-bound droplets at both zero and finite temperatures
\cite{Boudj1,Boudj2, Boudj3,Boudj18}.
An essential feature of the variational TDHFB theory is that it takes into account the normal and anomalous fluctuations which are crucial, in order to have a consistent description of the droplet.
Remarkably, in 3D our theory shows an excellent agreement with DMC data and the previous theoretical results for the energy and the equilibrium density \cite{Boudj18}.  
Regarding self-bound droplets of single dipolar BECs, the TDHFB provides also satisfactory explanations to experimental results 
and gives best match with the latest QMC simulation \cite{Boudj1}. 

In this paper, we investigate many-body effects and impacts of higher-order quantum fluctuations on the ground-state properties
of self-bound droplets of 2D symmetric Bose mixtures at both zero and finite temperatures using our HFB theory.
At finite temperature this exotic states of matter remains largely unexplored most likely due to  the self-evaporation i.e.
non existence of collective excitations below the particle-emission threshold \cite{Petrov}.
We calculate analytically the contribution to the sound velocity, the ground-state energy and the free energy from higher-order quantum and thermal fluctuations.
At zero temperature, the energy has a minimum at a finite density corresponding to a self-bound liquid-like droplet state. 
The obtained ground-state energy shows an excellent concordance with the  DMC results of Ref.\cite{Astr}, 
indicating the relevance of our model. We analyze also the behavior of  the depletion and the anomalous correlations of the droplet in terms of the equilibrium density.
At finite temperature, we find that the self-bound droplet may occur only at a certain critical temperature well below the Berezinskii-Kosterlitz-Thouless (BKT) transition due 
to the crucial role played by thermal fluctuations effects. Such a critical temperature decreases as the strength of interspecies interactions grows.
Furthermore, we  show that our formalism provides an extended finite-temperature GPE in which higher-order logarithmic factors are added to the nonlinear term of the condensate.
We use this model and discuss in particular the role of the quantum fluctuations play in the density profiles and the width of the droplet.
To the best of our knowledge this is the first theoretical investigation of 2D self-bound Bose mixtures at finite temperature in the presence of higher-order corrections.

\section*{Results}

\section*{Fluctuations and thermodynamics of 2D Bose mixtures} \label{FTh}

We consider a weakly interacting 2D Bose mixture with equal masses. 
The dynamics of this system including the effect of quantum and thermal fluctuations is governed by the coupled TDHDB equations which can be written in compact form as \cite{Boudj2,Boudj3,Boudj18,Boudj4}:
\begin{subequations}\label{TDHFB}
\begin{align}
&i\hbar \frac{d  \Phi_j}{d t}  = \bigg (h_j^{sp}+ g_j n_j+ g_{12} n_{3-j}+ \delta \mu_{j\text{LHY}} \bigg)\Phi_j,  \label{TDHFB1}\\
&i\hbar \frac{d \rho_j}{d t} =-2\left[\rho_j, \frac{d{\cal E}}{d\rho_j} \right], \label{TDHFB3} 
\end{align}
\end{subequations}
where $\rho_j ({\bf r},t)$ is the single particle density matrix of a thermal component  defined as
$$
\rho_j=\begin{pmatrix} 
\langle \hat{\bar{\psi}}^\dagger\hat{\bar{\psi}}\rangle & -\langle\hat{\bar{\psi}}\hat{\bar{\psi}}\rangle\\
\langle\hat{\bar{\psi}}^\dagger\hat{\bar{\psi}}^\dagger\rangle& -\langle\hat{\bar{\psi}}\hat{\bar{\psi}}^\dagger\rangle
\end{pmatrix}_j,
$$
and
${\cal E} = \sum_{j=1}^2 \bigg[ \int d{\bf r} \, \left( \Phi_j^*  h_j^{sp} \Phi_j + \hat{\bar \psi}_j^\dagger  h_j^{sp}  \hat{\bar \psi}_j +g_j  n_j^2/2 \right) \bigg] 
+ g_{12} \int d{\bf r}  n_1n_2 + {\cal E}_{\text {LHY}}$, is the energy of the system with 
${\cal E}_{\text{LHY}} = \sum_{j=1}^2 (g_j/2) \int d{\bf r} \big( 2\tilde n_j n_j-\tilde n_j^2 +|\tilde m_j|^2 + \tilde m_j^*\Phi_j^2+ \tilde m_j {\Phi_j^*}^2 \big)$
being the LHY correction to the energy.
In Eqs.(\ref{TDHFB}) $h_j^{sp} =-(\hbar^2 /2m_j) \Delta-\mu_j$ is the single particle Hamiltonian, $\mu_j$ is the chemical potential of each component, 
$\delta \mu_{j\text{LHY}} ({\bf r}) \Phi_j({\bf r})=g_j \big[\tilde n_j ({\bf r})\Phi_j({\bf r}) +\tilde m_j ({\bf r})\Phi_j^*({\bf r})\big]$ is the relevant  LHY term which is obtained self-consistently,
$\hat{\bar \psi}_j({\bf r})=\hat\psi_j({\bf r})- \Phi_j({\bf r})$ is the noncondensed part of the field operator with $\Phi_j({\bf r})=\langle\hat\psi_j({\bf r})\rangle$, 
$n_{cj}=|\Phi_j|^2$ is the condensed density,  $\tilde n_j= \langle \hat{\bar{\psi}}_j^\dagger\hat{\bar{\psi}}_j\rangle$ is the noncondensed density, 
$\tilde m_j=\langle\hat{\bar{\psi}}_j\hat{\bar{\psi}}_j\rangle$ is the anomalous correlation, and $n_j=n_{cj}+\tilde n_j$ is the total density of each species.
In 2D Bose gases, the intra- and interspecies coupling strengths are given by $g_j=4\pi \hbar^2/\left[m\ln \left(4e^{-2\gamma}/ a_j^2 \kappa^2\right)\right]$, 
and $g_{12}=g_{21}=4\pi \hbar^2/ \left[m\ln \left(4^2e^{-2\gamma}/ a_{12}^2\kappa^2\right)\right]$,
where $a_j$ and  $a_{12}$ being the 2D scattering lengths among the particles (see, e.g., \cite{Astr, Pop, Boudj7}), 
$\gamma=0.5772$ is Euler's constant.  An adequate value of the cutoff $\kappa$ can be obtained in the weakly interacting regime.
In such a case, attraction (repulsion) can be reached when the scattering lengths are exponentially large (small) compared to the mean interparticle separation \cite{Astr}.\\

The presence of the noncondensed and anomalous densities in Eqs.(\ref{TDHFB}) enables us to derive higher-order quantum corrections without any ad-hoc assumptions 
in contrast to the standard GPE. In our formalism $\tilde n_j$ and $\tilde m_j$ are related with each other via
\begin{equation}  \label{Inv1}
I_j= (2\tilde n_j+1)^2- 4|\tilde m_j |^2.
\end{equation}
This equation which steems from the conservation of the Von Neumann entropy, represents the variance of the number of noncondensed particles \cite{Boudj6,Boudj7}. 
Equation (\ref{Inv1}) clearly shows that the anomalous density is not negligible even at zero temperature ($I \rightarrow 1$), 
contrary to what has been argued in the literature. Hence,  $\tilde m$  is crucial for the stability of Bose gases.  
Its involvement in such systems leads to a double counting of the interaction effects \cite{Boudj3}.

In order to calculate the elementary excitations and fluctuations of a homogeneous Bose mixture, we linearize Eqs.(\ref{TDHFB}) using the
generalized random-phase approximation (RPA): 
$\Phi_j = \sqrt{n_{cj}}+\delta \Phi_j $, $\tilde n_j=\tilde n_j+\delta \tilde n_j$,  and $\tilde m_j=\tilde m_j+\delta \tilde m_j$, 
where $\delta \Phi_j ({\bf r},t)= u_{jk}  e^{i {\bf k \cdot r}-i\varepsilon_k t/\hbar}+v_{jk} e^{i {\bf k \cdot r}+i\varepsilon_k t/\hbar} \ll \sqrt{n_{cj}}$, 
$\delta \tilde n_j \ll \tilde n_j$, and $\delta \tilde m_j \ll \tilde m_j$ \cite{Boudj2,Boudj3}. 
Since we restrict ourselves to second-order in the coupling constants, we keep only the terms which describe the coupling to the condensate and 
neglect all terms associated with fluctuations $\delta \tilde n_j$ and $\delta \tilde m_j$ (see Methods).
The obtained second-order coupled TDHFB-de Gennes equations which are similar to the Beliaev's equations \cite{Gu,Beleav,Beleav1}, 
provide correction terms to the Bogoliubov formula for the energy spectrum:
$\varepsilon_{k\pm}= \sqrt{E_k^2+2E_k \mu_\pm}$ \cite{Boudj3}, where $\mu_{\pm}=  \bar g_1 n_{c1} [1 + \alpha \pm \sqrt{ (1-\alpha)^2 +4 \Delta ^{-1}\alpha }]/2$,
$\Delta=\bar g_1\bar g_2/g_{12}^2$, and $\alpha=\bar g_2 n_{c2}/\bar g_1 n_{c1}$. 
Here the density-dependent coupling constants  $\bar g_j=g_j(1+\tilde m_j/n_{cj})$ have been introduced in order to reinstate the gaplessness of the spectrum \cite{Boudj3}.

The noncondensed and anomalous densities can be computed through Eq.(\ref{Inv1}) \cite{Boudj3,Boudj18}
\begin{equation}\label {nor}
\tilde n_{\pm}=\frac{1}{2}\int \frac{d \bf k} {(2\pi)^2} \left[\frac{E_k+ \mu_{\pm}} {\varepsilon_{k \pm}} \sqrt{I_{k \pm}}-1\right],
\end{equation}
and
\begin{equation}\label {anom}
\tilde m_{\pm}=-\frac{1}{2}\int \frac{d \bf k} {(2\pi)^2} \frac{ \mu_{\pm} } {\varepsilon_{k \pm}} \sqrt{I_{k\pm}},
\end{equation}
where $I_{k\pm} =\text {coth} ^2\left(\varepsilon_{k\pm}/2T\right)$ \cite{Boudj3,Boudj18}.

At $T=0$, integral (\ref {nor}) gives the following expression for the total  depletion $\tilde n=\tilde n_++\tilde n_-$:
\begin{equation}\label {nor1}
\tilde n= \frac{m^2}{4\pi \hbar^2} \sum_{\pm} c_{s\pm}^2,
\end{equation}
where $c_{s\pm}^2= \mu_{\pm}/m$ are the sound velocities which can be evaluated selfconsistently.\\

The integral in Eq.(\ref{anom}) is ultraviolet divergent and necessitates to be regularized \cite{Boudj7, Anders, Luca}.
We use the dimensional regularization that is asymptotically accurate for weak interactions \cite{Boudj7,Yuk}.
Then one analytically continues the result to finite coupling including a low-energy cutoff  $\epsilon_c= \hbar^2 \kappa^2 /m \gg \mu_{\pm}$ \cite{Boudj7, Anders, Luca}.
This yields for the total anomalous density $\tilde m=\tilde m_++\tilde m_-$:
\begin{align}\label {anom1} 
\tilde m= \frac{m^2}{4\pi \hbar^2} \sum_{\pm} c_{s\pm}^2 \ln \bigg(\frac{m c_{s\pm}^2} { \epsilon_c} \bigg), 
\end{align}
For $g_{12}=0$, Eqs.(\ref {nor1}) and (\ref{anom1}) recover those obtained by our second-order TDHFB-de Gennes equations \cite{Boudj7} for a single component condensate.

The knowledge of the noncondensed and anomalous densities allows one to predict higher-order corrections to the free energy. 
In the frame of our formalism, it can be written as:
\begin{equation}\label {fergy}
F=E+T \int \frac{d \bf k} {(2\pi)^2}\ln\left(\frac{2}{\sqrt{I_{k\pm}}+1}\right),
\end{equation}
where 
\begin{align}\label{gerny}
E=E_0+\frac{1}{2} \sum_{\pm} \int \frac{d \bf k} {(2\pi)^2} \left(\varepsilon_{k \pm} - E_k-\mu_{\pm} \right),
\end{align}
is the ground-state energy, and 
$E_0=\frac{1}{2}  \sum_{j=1}^2 g_j ( n_{cj}^2+ 4 n_{cj} \tilde n_j +2\tilde n_j^2 +\tilde m_j^2 + 2 n_{cj} \tilde m_j) +g_{12}  n_1 n_2$.
The second term in Eq.(\ref{gerny}) accounts for the LHY quantum corrections. It can be computed using the above dimensional regularization
where only the bound modes that have energy lower than the magnitude of the binding energy are included in the integral \cite{Hu}.
The subleading term in Eq.(\ref{fergy}) which represents the thermal effects is finite.
Gathering  quantum and thermal fluctuations contributions to the free energy (\ref{fergy}), we get
\begin{align}\label {fergy1}
F=E_0+\frac{m^3}{ 8\pi \hbar^2}\sum_{\pm} c_{s\pm}^4 \ln \left(\frac{\sqrt{e} m c_{s\pm}^2} {\epsilon_c}\right)-\sum_{\pm} \frac{\zeta (3)}{(\hbar c_{s\pm})^2}T^3,
\end{align}
here we employed the identity $\int_0^{\infty} dx x \ln[2/(\coth(x/2)+1)]=-\zeta (3)$, where $\zeta (3)$  is the Riemann zeta function.
Expression (\ref{fergy1}) extends naturally the results of Petrov and Astrakharchik \cite{Astr} since it takes into account both many-body and temperature effects.

It is worth stressing that Eqs.(\ref{nor})-(\ref{fergy1}) are self-consistent and must be solved iteratively.

\section*{Self-bound droplets} \label{SBD}

Now, we consider 2D symmetric Bose mixture with repulsive intraspecies interaction and attractive interspecies interaction where $a_{12}^{-1}\ll \sqrt{n} \ll a^{-1}$.
The atoms are chosen to have equal intra-component scattering lengths $a_1=a_2 =a$ and equal atom densities $n_1=n_2= n$, $\tilde n_1=\tilde n_2=\tilde n$, 
and $\tilde m_1=\tilde m_2=\tilde m$. For the sake of simplicity we put $\hbar=m=1$.

\begin{figure}
\centerline{
\includegraphics[scale=0.8]{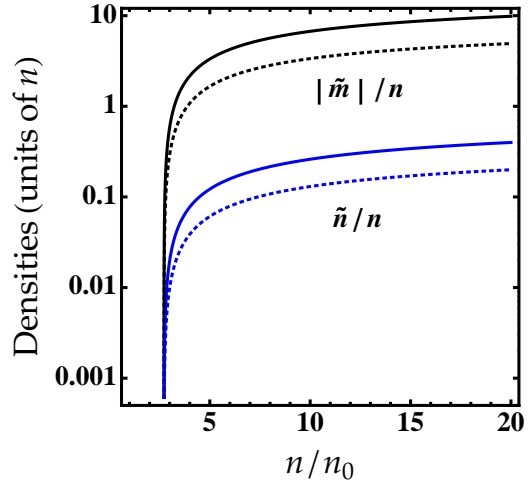}}
 \caption{Noncondensed (\ref{nor2}) and anomalous (\ref{anom2}) fractions at the equilibrium as a function of $n/n_0$ for different values of $\ln(a_{12}/a)$.
Solid lines: $\ln(a_{12}/a)=5$. Dotted lines: $\ln(a_{12}/a)=10$.}
\label{AD} 
\end{figure}

\subsection*{Zero-temperature case} 

At zero temperature,  the properties of self-bound ultradilute Bose mixtures can be analyzed by minimizing the ground-state energy with respect to the density 
or equivalently using the zero-pressure condition $P=\mu n-E/S=0$, where $S$ is the surface area \cite{Astr}.
According to the method outlined in Ref.\cite{Astr}, we introduce a new set of coupling constants given as: 
$g=4\pi/\ln \left(4e^{-2\gamma}/ a^2  \epsilon_0 \right)$ and $g_{12}=4\pi/\ln \left(4e^{-2\gamma}/ a_{12}^2  \epsilon_0 \right)$,
where $\epsilon_0=4e^{-2\gamma}/ a_{12} a$ has been choosed in such a way that the condition $g^2=g_{12}^2$ must be fulfilled. 
This implies that $c_{s-}=0$ which means that $\tilde n_-=\tilde m_-=0$.
Then, the corrected sound velocity can be obtained via  $c_s^2= 2gn \big(1-\tilde n/2n+\tilde m/2n\big)$ 
(here we set $c_{s+}=c_s$, $\tilde n_+=\tilde n$, and $\tilde m_+=\tilde m$ for convenience).
For the purpose of analytical tractability, we keep only lowest order in $\tilde n$ and $\tilde m$. This gives: 
\begin{equation}\label {SV1}
\frac{c_s^2}{c_{s0}^2} = \frac{1}{4\pi} \left(\frac{n}{n_0}\right) \bigg[\ln \left(\frac{n}{n_0}\right)-1\bigg],
\end{equation}
where $n_0 = \epsilon_c/\big(2g e^{8\pi}\big)$ is the equilibrium density which can be obtained by minimizing  the ground-state energy (\ref{fergy1}) with respect to the density.
The sound velocity at the equilibrium is defined as $c_{s0}^2=gn_0$. 
Cearly,  Eq.(\ref{SV1}) predicts an imaginary sound velocity which may lead to a complex energy functional.
Similar behavior has been reported in \cite{Ota,Gu} for 3D droplets. In the Petrov's work \cite{Petrov,Astr}, such a dynamically unstable phonon mode has been completely ignored
under the assumption that its contribution is negligibly small. To stabilize the sound velocity and obtain the associated ground-state energy, 
we should include higher-order fluctuations (see below).

The noncondensed and anomalous densities of the droplet corresponding to the sound velocity (\ref{SV1}) read:
\begin{align}\label {nor2}
\frac{\tilde n_{\text{eq}}}{n}=\frac{\ln (n/n_0)-1}{\ln\left (a_{12}/a\right)}, 
\end{align}
and
\begin{align}\label {anom2} 
\frac{\tilde m_{\text{eq}}}{n}=\frac{\tilde n}{n}  \ln \bigg[\frac{n/n_0 (\ln (n/n_0)-1)}  {8\pi e^{8\pi}}\bigg], 
\end{align}
Figure \ref{AD} shows that  $\tilde m$ is larger than $\tilde n$ regardless of the value of  $\ln(a_{12}/a)$ as in the case of self-bound droplets in 3D Bose mixtures \cite{Boudj18}.
Both densities are increasing with decreasing  $\ln(a_{12}/a)$.

\begin{figure}
\centerline{
\includegraphics[scale=0.46]{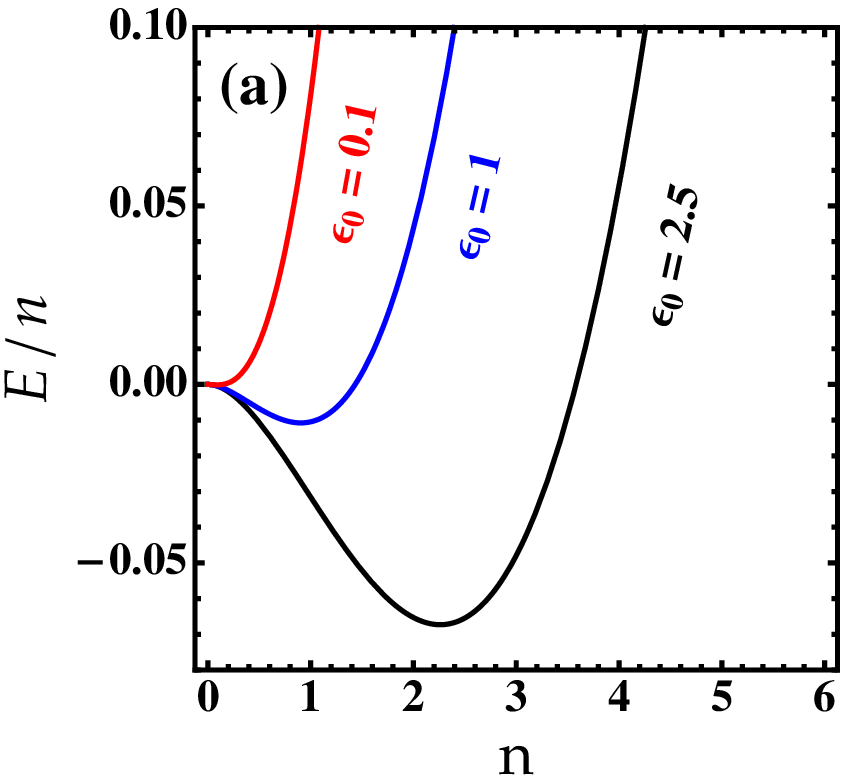}
\includegraphics[scale=0.45]{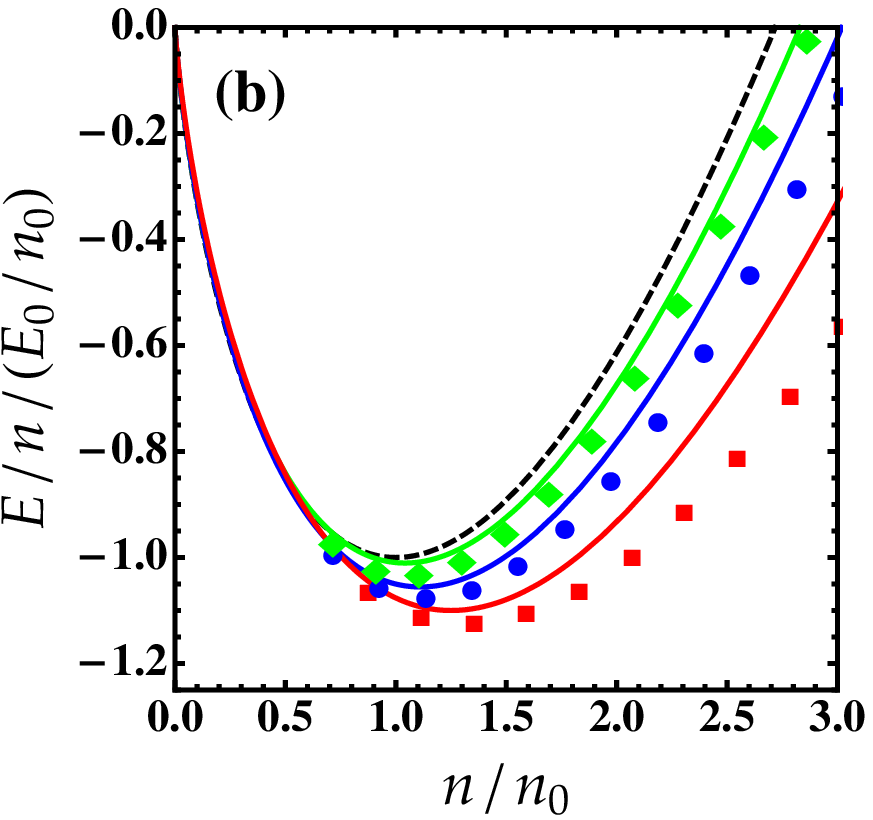}}
 \caption{(a) The ground-state energy $E/n$  from Eq.(\ref{gerny})  for several values of $\epsilon_0$ and $g = 0.45$.
(b) The ground-state energy as a function of $n/n_0$.
Solid lines correspond to our beyond-LHY results up to second-order in $\tilde n$ and $\tilde m$. Dashed line corresponds to the Bogoliubov theory \cite{Astr}. 
Green diamonds ($\ln (a_{12}/a)=20$), blue circles ($\ln (a_{12}/a)=10$), and red squares ($\ln (a_{12}/a)=5$)  correspond to the DMC data of \cite{Astr}. 
Here $E_0=E(n_0)$.}
\label{GSE} 
\end{figure}

Let us now calculate the ground-state energy for 2D symmetric Bose mixtures by seeking the effect of higher-order fluctuations where a numerical method is used
to treat the involved integration.  The results are depicted in Fig.\ref{GSE}.

We see from Fig.\ref{GSE}.(a) that the variation of the energy-cutoff which depends on interspecies interactions may strongly change the position of the local minimum of the energy 
leading to affect the stability and the existence of the droplet. For instance, for $\epsilon_0 \leq 0.1$, the local minimum disappears  and the energy becomes positive
indicating that the droplet may turn into a soliton-like many-body bound state in good agreement with the predictions of Refs.\cite{Hammer,Chiq,Hu}.

In Fig.\ref{GSE}.(b) we compare our results for the ground-state energy up to second order in $\tilde m$ and $\tilde n$  of the iteration method
with the DMC data and the Bogoliubov theory \cite{Astr}. 
We see that when $\ln(a_{12}/a)$ gets larger, our results excellently agree with the DMC simulations and improve the standard Bogoliubov findings.
This implies that for large $\ln(a_{12}/a)$, the HFB predictions become increasingly accurate due to the considerable role of higher-order terms 
arising from the normal and anomalous fluctuations. 
Our results diverge from the DMC simulations only for very small values of interspecies interaction $\ln(a_{12}/a)<5$ and higher densities.

\subsection*{Finite-temperature case} 

In homogeneous 2D Bose gases, thermal fluctuations are strong enough to prohibit the formation of a true BEC at any nonzero temperature \cite{merm, hoh}.
However, according to BKT \cite{Berz,KT},  quasicondensate (or a condensate with only local phase coherence) takes place below the BKT transition temperature. 
The transition from a noncondensed state to quasicondensate occurs through the formation of bound vortex-antivortex pairs \cite{Prok,Chom1}.
In such a quasicondensate, the phase coherence governs only regime of a size smaller than the size of the condensate, marked by the coherence length $l_{\phi}$ \cite {GPS}.
Therefore, below the BKT transition temperature one can use the HFB theory to describe the true BEC \cite{Boudj7,Boudj17, Dalf} 
even though it cannot predict the critical fluctuations near the BKT region. 

At finite temperature, the free energy becomes  divergent since $c_{s-}=0$ results in an unstable droplet in contrast to the zero-temperature case.
Hence, to properly study the finite-temperature behavior of the 2D self-bound droplet, the sound velocity must be finite:
\begin{equation}\label {SV2}
c_{s\pm}^2=  \delta g_{\pm} n\bigg [1+\frac{\delta g_{\pm}}{4\pi}\ln \left(\frac{n\delta g_{\pm} }{\epsilon_c e}\right) \bigg],
\end{equation}
where $\delta g_{\pm}=g\pm g_{12}$.

\begin{figure}
\centerline{
\includegraphics[scale=0.8]{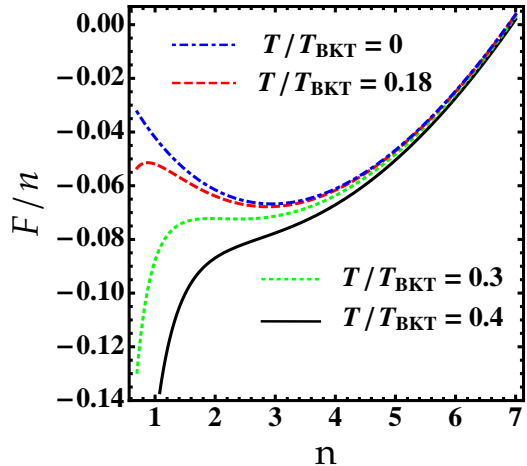}}
 \caption{ The free energy $F/n$ as a function of the density for different values of temperature, $T/T_{\text{BKT}}$.
Parameters are: $g= 0.45$, $g_{12}=0.2$ and $\epsilon_0= 4.8$}.
\label{FE} 
\end{figure}

Minimizing the resulting free energy, we could observe the equilibrium emergence of the droplet, as visible in Fig.\ref{FE}.
The temperature is normalized to $T_{\text{BKT}}$ which is defined for a symmetric mixture according to $T_{\text{BKT}}=\pi N \ln(380/g)/S$ \cite{Dalf}.
It has been demonstrated that the interspecies interaction plays a  minor role near the BKT critical temperature \cite{Kar, Kob}.
We see that the free energy $F$ diverges like $n^{-1}$ as the density goes to zero due to the presence of thermal fluctuations effects.
Well below the BKT transition i.e. $0 <T  \lesssim 0.3\,T_{\text{BKT}}$, $F$ develops a local maximum which corresponds to an unstable droplet,
and a local minimum supporting a higher density stable self-bound solution.
In such a regime, thermally excited atoms that occupy continuum modes are unbound and leave the droplet result in 
a process of self-cooling predicted earlier by Petrov  \cite{Petrov}.
The two solutions disappear at the critical temperature ($T=T_c\simeq 0.3\,T_{\text{BKT}}$) revealing that the liquid-like droplet start to evaporate.
Increasing further the  temperature ($T > T_c$), the free energy increases without any special structure and thus, 
the self-bound state loses its peculiar self-evaporation phenomenon and entirely destroys eventually. 
The same situation takes place for dipolar droplets in a single BEC \cite{Boudj16, Boudj15, Ayb}  and in dual condensates \cite{Boudj2,Boudj17}. 
Note that $T_c$ strongly relies on $\epsilon_0$ and hence, on the interspecies interactions as we shall see below.

\begin{figure}
\centerline{
\includegraphics[scale=0.8]{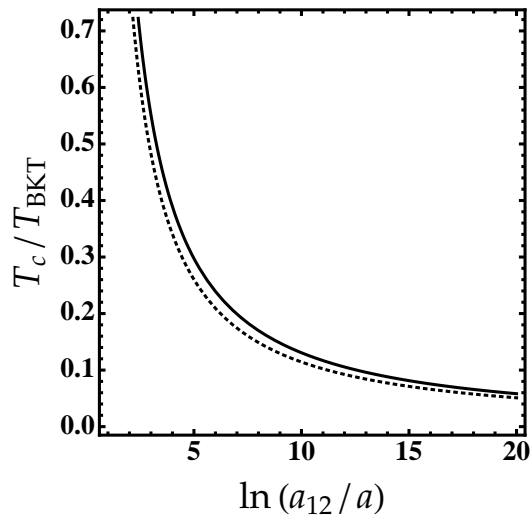}}
 \caption{ Critical temperature normalized to $T_{\text{BKT}}$ as a function of $\ln (a_{12}/a)$.
Solid line: without higher-order effects. Dotted line: higher-order effects.}
\label{CT} 
\end{figure}

The critical temperature above which the BEC-droplet phase transition occurs can be determined by minimizing the free energy.
For $\delta g_{-} \ll \delta g_{+}$, one has 
\begin{equation}\label{gernyT}
\frac{T_c} {T_{\text{BKT}}}  \simeq \frac{\ln[(n/2e^{4\pi-2}n_0)-1]^{1/3} }  {(\pi\zeta (3))^{1/3}  \ln( a_{12}/a) \ln [\ln( a_{12}/a)95/\pi]},
\end{equation}
As shown in Fig.\ref{CT} for fixed density $n/n_0$, the droplet critical temperature decreases with the interspecies interaction $\ln (a_{12}/a)$
regardless the presence or not of the higher-order effects. For example for $\ln (a_{12}/a)=20$, the droplet reaches its thermal equilibrium at ultralow temperature 
($T_c \simeq 0.06 T_{\text{BKT}}$). We see also that higher-order corrections may reduce the critical temperture.

\section*{Generalized finite-temperature Gross-Pitaevskii equation} \label{FSE}

\begin{figure}
\centerline{
\includegraphics[scale=0.45]{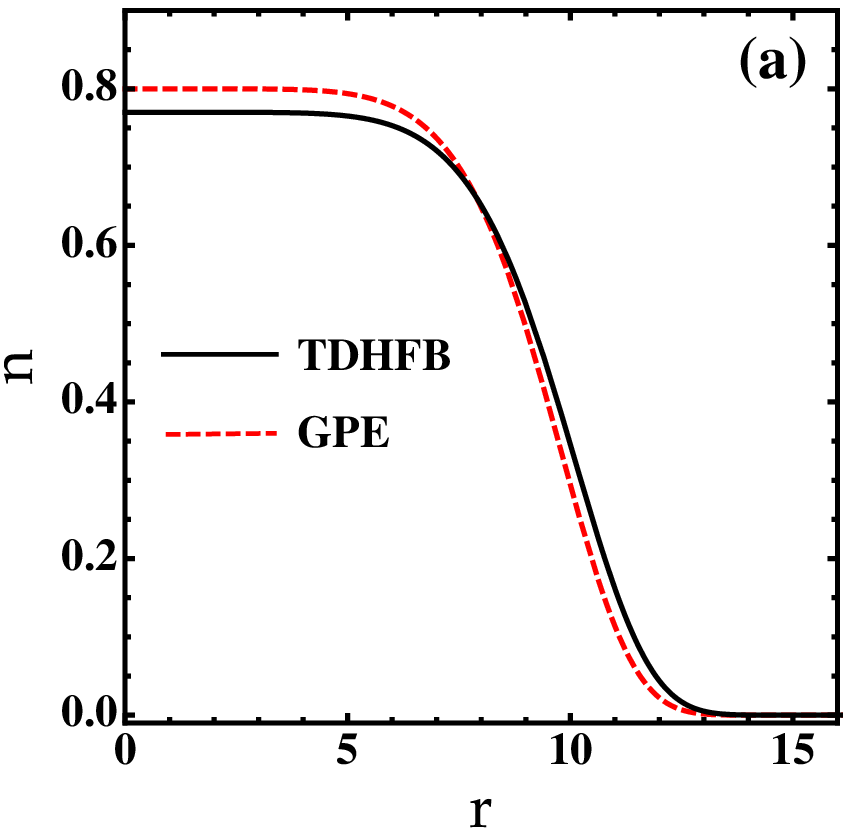}
\includegraphics[scale=0.45]{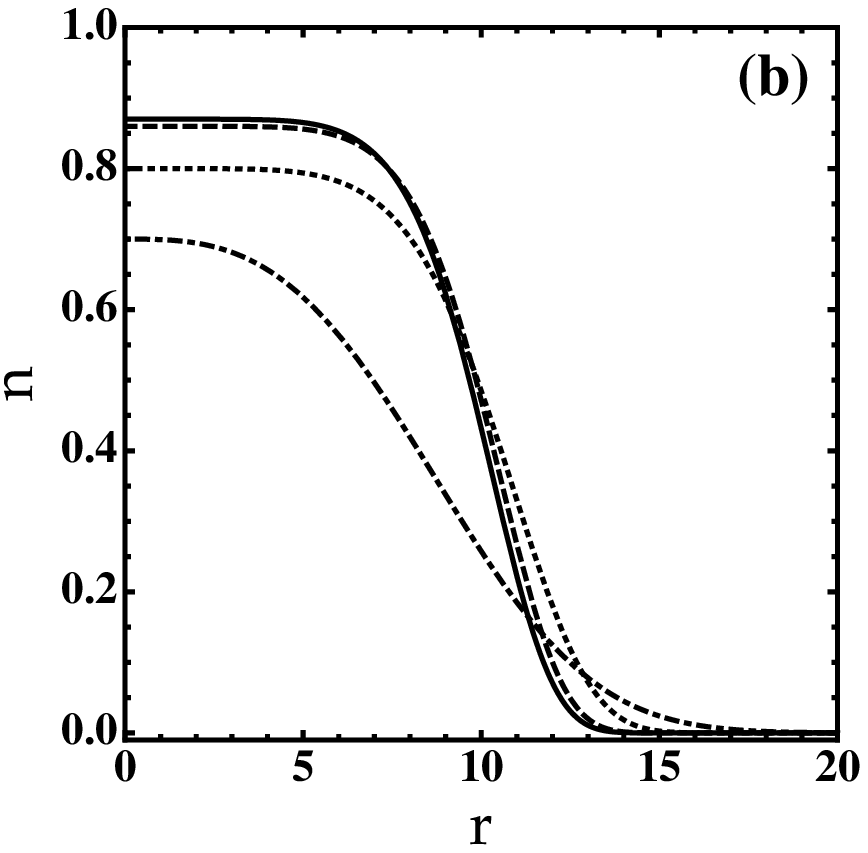}}
 \caption{ (a) Density profiles of the self-bound droplet obtained from the numerical solution of Eq.(\ref{TDH11}) at zero temperature for $N=1000$ atoms and $\ln (a_{12}/a)=20$.
(b) Density profiles of the self-bound droplet obtained at different values of temperature for $N=1000$ atoms and $\ln (a_{12}/a)=20$.
Solid line: $T=0$. Dashed line: $T=0.18T_{\text{BKT}}$. Dotted line: $T=0.3T_{\text{BKT}}$. Dotted-Dashed line: $T=0.4T_{\text{BKT}}$.}
\label{DP} 
\end{figure}

\begin{figure}
\centerline{
\includegraphics[scale=0.8]{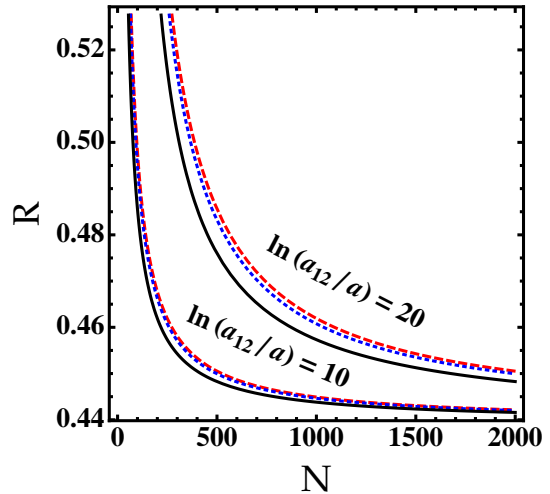}}
 \caption{ The self-bound droplet width as a function of the particles number $N$ for different values of $\ln (a_{12}/a)$.
The solid line corresponds to our generalized GPE (\ref{TDH11}). The red dashed line corresponds to the standard GPE \cite{Astr}. 
The blue dotted line corresponds to the variational calculation.}
\label{DS} 
\end{figure}

In this section, we consider the finite size effects on equilibrium properties of the self-bound droplet. 
The basic idea behind finite size contributions to the droplet's energy is that the quantities $\Phi$, $\tilde n$, and $\tilde m$ must vary slowly at the scale of the extended healing length. 
As a consequence, we can include higher-order corrections locally as nonlinear terms in the TDHFB equations and treat them classically.
For simplicity, we will ignore the dynamics of the thermal cloud and the anomalous correlations. 
Therefore, the TDHFB equation (\ref{TDHFB1}) leads directly to the generalized finite-temperature GPE
\begin{align} \label {TDH11}
i\frac{d  \Phi}{d t} = -\frac{\nabla^2}{2} \Phi+ \bigg[\frac{8\pi}{\ln^2(a_{12}/a)} \ln \left(\frac{|\Phi|^2}{\sqrt{e}n_0}\right) \alpha +\alpha _T \bigg]|\Phi|^2 \Phi, 
\end{align}
where $\alpha \simeq \ln \big[(e n/n_0)\ln (n/en_0)/8\pi e^{8\pi}\big]$, and 
$\alpha_T\simeq \zeta (3) T^3\ln(a_{12}/a) /\left[2\ln \left(|\Phi|^2/e ^{4\pi+1}n_0\right) |\Phi|^4\right] [1/\ln \left(|\Phi|^2/e ^{4\pi+1}n_0\right) +1]$.
Importantly, the generalized finite-temperature GPE (\ref{TDH11}) extends naturally the GPE of Ref.\cite{Astr} since it takes into account higher-order quantum and thermal corrections.

The stationary  solutions of Eq.(\ref{TDH11}) can be found via the transformation $ \Phi({\bf r}, t)= \Phi({\bf r}) \exp( {-i\mu t})$.
We solve the resulting static equation numerically using the split-step Fourier transform \cite{Sem}. 
In Fig.\ref{DP}, we plot the density profiles as a function of the radial distance at both zero and finite temperatures.
As can be seen in Fig.\ref{DP}.(a), the density $n$ is flattened in accordance to the liquid character of the condensate.    
The obtained density is compared with the predictions of the GPE-LHY theory \cite{Astr}.
Our results show a slight deviation downwards for distance $r<8$ with respect to the findings of Ref.\cite{Astr} owing to the higher-order quantum fluctuations.
At temperatures $T \lesssim T_c$, the droplet exhibits a weak-temperature dependence [see Fig.\ref{DP}.(b)]. Whereas, at $T \geq T_c$, the droplet has  a Gaussian-like shape pointing out that
the system experiences droplet-BEC phase transition.

To evaluate the width of the self-bound droplet, we first use the following trial wavefunction : $\Phi(r)= \exp ({-r^2/2R^2})/\sqrt{\pi R^2}$, 
where $R$ is the self-bound droplet width. Then,  we minimize the resulting functional energy with respect to $R$.
In the absence of the higher-order corrections, the width takes the form 
\begin{equation} \label {Width}
R\simeq  \frac{1}{\sqrt{\pi}} \exp{[-(1/4) + \ln^2 (a_{12}/a)/8N]}.
\end{equation}
This analytical prediction is reported in Fig.\ref{DS} and compared with the numerical results of our generalized GPE (\ref{TDH11}). 
We see that the width of the droplet decreases exponentially versus the number of particles. The interspecies interactions $a_{12}/a$ lead also to reduce the width.
The comparison between our predictions and those of Petrov \cite{Astr} indicates that the higher-order quantum effects may shift the droplet width.
At finite temperature one can expect that the droplet size increases significantly only at temperatures $T\gtrsim T_c$.  
Above such a temperature the self-bound droplet is in its thermal stabilization.

\section*{Discussion}

We studied the equilibrium properties of symmetric self-bound droplets of 2D binary BEC beyond the standard LHY treatment, 
at both zero and finite temperatures.
We computed higher-order corrections to the excitations spectrum,  the sound velocity, the normal and anomalous correlations, and the free energy.
These corrections improve the ground-state energy obtained from the Bogoliubov approach \cite{Astr} predicting an energy in good agreement with recent DMC simulations
owing to the non-negligeable role of higher order terms.
At finite temperature, we revealed that the droplet occurs at temperature well below the BKT transition and destroys when the temperature 
becomes slightly larger than the ground-state energy of the droplet due to the thermal fluctuations effects. We found that the interspecies interaction tends to lower the critical temperature.
We analyzed in addition the finite-size droplets in the framework of our generalized finite-temperature GPE.
As outlined above, one can infer that in 2D mixtures, the droplet survives only in an ultradilute regime and at ultralow temperatures.

Our results could be extended in weakly interacting quasi-2D Bose mixtures as long as the following condition is fulfilled 
$0<-a_{12}^{3D}<a^{3D}\ll l_0$ \cite{Astr},
where $a^{3D}$ and $a_{12}^{3D}$ are the 3D intra and interspecies scattering lengths, and $l_0$  is the oscillator length in the confinement direction.
The creation of such 2D mixture droplets in the experiment, still remains a challenging question.

\section*{Methods}

\subsection*{Derivation of the condensate fluctuations}

As we concluded in the main text, for a thermal distribution at equilibrium and by working in the momentum space, one has \cite{Boudj7}
\begin{equation} \label {TDH1}
\rho_{mn}({\bf r}-{\bf r'})= \int \frac{d {\bf k}} {(2\pi)^3} e^{i{\bf k}. ({\bf r}-{\bf r'}) } \rho_{mn} ({\bf k}), 
\end{equation}
where $ \rho_{mn} ({\bf k})$ is the Fourier transform of $\rho_{mn}({\bf r}-{\bf r'})$. After some algebra,  expression (\ref{Inv1}) turns out to be given as:
\begin{equation}\label {heis}
I_{k\pm}= (2\tilde{n}_{\pm k}+1)^2-|2\tilde{m}_{\pm k}|^2= \coth^2\left(\varepsilon_{k\pm}/2T\right),
\end{equation}
From Eq.(\ref{heis}) we can straightforwardly derive the expressions (\ref{nor}) and (\ref{anom})  describing the normal and anomalous correlations.
For an ideal Bose gas where the anomalous density vanishes,  $I_{k} =\text {coth} ^2\left(E_{k}/2T\right)$ \cite{Boudj3}.

\subsection*{TDHFB-de Gennes equations}

We use the generalized RPA which consists of imposing small fluctuations of the condensates, the noncondensates, and the anomalous components, respectively, as: 
$\Phi_j = \sqrt{n_{cj}}+\delta \Phi_j $, $\tilde n_j=\tilde n_j+\delta \tilde n_j$,  and $\tilde m_j=\tilde m_j+\delta \tilde m_j$,  
where $\delta \Phi_j \ll \sqrt{n_{cj}}$, $\delta \tilde n_j \ll \tilde n_j$, and $\delta \tilde m_j \ll \tilde m_j$ \cite{Boudj3}. 
We then obtain the TDHFB-RPA equations:
\begin{align} 
i\hbar \delta  \dot \Phi_j & = \left[ h_j^{sp}+ 2\bar g_j n_{cj}+2g_j \tilde n_j + g_{12} n_{3-j} \right] \delta \Phi_j  \label {RPA1} \\ 
    &+\bar g_j n_{cj} \delta \Phi_j^*+ 2g_j \sqrt{n_{cj}} \delta \tilde n_j  + g_{12}\sqrt{n_{c{3-j}}} \delta \tilde n_{3-j} \nonumber \\
&+ g_{12} \sqrt{ n_{cj} n_{c{3-j}} } (\delta \Phi_{3-j}+\delta \Phi_{3-j}^*),  \nonumber 
\end{align}
and 
\begin{align} 
i\hbar \delta  \dot{\tilde m}_j &=  4\left[ h_j^{sp}+2g_j n_j+g_j \bar g_j /4(\bar g_j-g _j) (2\tilde n_j +1)+g_{12} n_{3-j} \right] \delta\tilde m_j   \label {RPA2} \\ 
&+ 8g_j \tilde m_j \left[ \sqrt{ n_{cj}}  (\delta \Phi_j+ \delta \Phi_j^*)+ \delta \tilde n_j+  \bar g_j /4(\bar g_j-g _j)\delta \tilde n_j \right] \nonumber\\
&+ g_{12}\tilde m_j \left[\sqrt{ n_{c{3-j}}}  (\delta \Phi_{3-j}+\delta \Phi_{3-j}^*) +\delta \tilde n_{3-j} \right], \nonumber
\end{align}
Remarkably, this set of equations contains a class of terms beyond second order. 
Note that we keep in Eqs.(\ref{RPA1})  and (\ref{RPA2}) only the terms which describe the coupling to the condensate and neglect all terms associated 
with $\delta \tilde n$ and $\delta \tilde m$ owing to the fact that we restrict ourselves to second-order in the coupling constants.

Inserting the transformation $\delta \Phi_j ({\bf r},t)= u_{jk}  e^{i {\bf k \cdot r}-i\varepsilon_k t/\hbar}+v_{jk} e^{i {\bf k \cdot r}+i\varepsilon_k t/\hbar}$ into Eqs.(\ref{RPA1}),  
we find the second-order coupled TDHFB-de Gennes equations for the quasiparticle amplitudes $u_{kj}$ and $v_{kj}$ :
\begin{equation} \label{BdG}
\begin{pmatrix} 
{\cal L}_1 & {\cal M}_1 &  {\cal A}  &  {\cal A}
\\
  {\cal M}_1 & {\cal  L}_1 &  {\cal A} &  {\cal A}
\\
  {\cal A} &  {\cal A}& {\cal L}_2 &  {\cal M}_2
\\
 {\cal A} &  {\cal A} &  {\cal M}_2 & {\cal  L}_2
\end{pmatrix}\begin{pmatrix} 
u_{1k} \\ v_{1k} \\ u_{2k}  \\ v_{2k} 
\end{pmatrix}=\varepsilon_k \begin{pmatrix} 
u_{1k} \\ -v_{1k}  \\ u_{2k}  \\ -v_{2k}
\end{pmatrix},
\end{equation} 
where $\int d {\bf r} [u_j^2( {\bf r})- v_j^2({\bf r})]=1$, 
${\cal L}_j = E_k+ 2 \bar g_j n_{cj}+ 2 g_j \tilde n_j + g_{12} n_{3-j} -\mu_j$,  ${\cal M}_j= \bar g_j n_{cj}$, and ${\cal A}=g_{12}\sqrt{n_{c1} n_{c2} }$.
Equations (\ref{BdG})  are appealing since they enable us to calculate in a simpler manner corrections to the excitations spectrum $\varepsilon_{k\pm}$ 
of homogeneous Bose mixtures (see the main text).


\section*{Acknowledgements}

We are grateful to Dmitry Petrov  and  Grigori Astrakharchik for sharing their DMC data and for useful comments. 
We acknowledge support from the Algerian Ministry of Higher Education and Scientific Research under Research Grant No. PRFU-B00L02UN020120190001.

\section*{Author contributions statement}
AB conceived the work, obtained the results, wrote and reviewed the manuscript.

\section*{Additional Information}
\subsection*{Competing financial interests:} 
The author declares no competing financial interests.

\end{document}